\documentclass[pre,amsfonts,showpacs,unsortedaddress]{revtex4}
\usepackage[english]{babel}
\usepackage[latin1]{inputenc}
\usepackage[dvips]{graphicx}
\usepackage{amsmath}
\usepackage{amssymb}
\usepackage{epsfig}
\usepackage{array}

\begin{document}

\title{Localization of shocks in driven diffusive systems
  without particle number conservation} 
\author{V. Popkov $^{1}$,
A. R\'akos$^{1}$, R. D. Willmann$^{1}$, A. B. Kolomeisky$^{2}$ and
G. M. Sch\"utz$^{1}$}
\affiliation{$^{1}$Institut f\"ur Festk\"orperforschung, Forschungszentrum J\"ulich - 
52425 J\"ulich, Germany}
\affiliation{$^{2}$
Department of Chemistry, Rice University, Houston, Texas 77005-1892
}

\date{\today}

\begin{abstract}
We study the formation of localized shocks in one-dimensional
driven diffusive systems with spacially homogeneous  
 creation and annihilation of particles
(Langmuir kinetics).We show how to obtain 
hydrodynamic equations which describe the density
profile in systems with uncorrelated steady state
 as well as in those exhibiting correlations. As a special 
example of the latter case the Katz-Lebowitz-Spohn model is considered.
The existence of a localized double density shock is demonstrated for the 
first time in one-dimensional driven diffusive systems. This corresponds
to phase separation into regimes of three distinct densities, separated
by localized domain walls. Our analytical approach is supported
by Monte-Carlo simulations.
\end{abstract}

\maketitle

\section{Introduction}
One-dimensional driven diffusive systems proved to be a rewarding
research topic in the past years \cite{Habil}.  
They were shown to exhibit boundary induced phase transitions
\cite{Krug}, spontaneous symmetry breaking \cite{Evans,Mukamel95} and phase
separation \cite{Mukamel1,Mukamel2}. 
Recently, attention turned to the case of systems without particle
conservation in the bulk. In Ref.\ \onlinecite{Kolomeisky2} the effect
of a single detachment site in the bulk of an asymmetric simple 
exclusion process (ASEP) was studied.
In Refs.\cite{Challet,Parmeggiani} the
interplay of the simplest one-dimensional driven model, 
the totally asymmetric exclusion process (TASEP) with local
absorption/desorption kinetics of single particles acting at all
sites, termed 'Langmuir kinetics' (LK) was considered. 
These models were inspired by
the dynamics of motor proteins\cite{motorprot4}
which move along cytoskeletal filaments 
in a certain prefered direction while detachment and attachment can
also occur between the cytoplasm and the filament,
and, in a very different setting, by dynamics of limit orders 
in a stock exchange market.
Being an equilibrium process, LK is well understood,
while the combined  process of TASEP and LK showed the new feature of
a localized shock in the density profile of the stationary state
\cite{Parmeggiani}.

The TASEP is defined on a one-dimensional lattice of size $L$.
Each site can either be empty or occupied by one particle.
 In the
bulk particles can hop from site $i$ to 
site $i+1$ with unit rate, provided the target site is empty. 
At site 1 particles can enter the lattice from a reservoir with
density $\rho_-$ provided the  
site is empty. They can leave the system
from site $L$ into a reservoir of density $\rho_+$ with rate $1-\rho_+$. 
Thus in the interior of the lattice the particle number is a conserved quantity.
The phase diagram and steady states of the TASEP as a function of the
boundary rates are known exactly  
\cite{Ligg75,Domany,ASEP}. Furthermore a theory of boundary induced phase
transitions exists, which explains
the phase diagram quantitatively in terms of the dynamics of shocks
\cite{Kolomeisky}. 
In the stationary state these shocks exist as an upward density shock
along the coexistence line between the high and low density phases,
i.e., they connect a region with low density to the left of the shock
position with a high density region to its right. The
shock performs symmetric random walk between the boundaries of 
the system.

\begin{figure}[h]
\begin{center}
\includegraphics[scale=0.7]{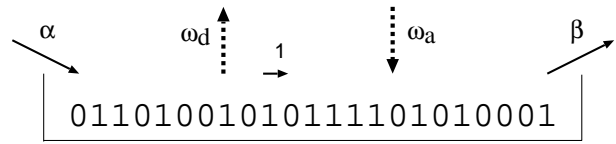}
\end{center}
\caption{Possible processes and their rates in the model of
the ASEP with Langmuir kinetics }
\label{model}
\end{figure}

\begin{figure}[h]
\begin{center}
\includegraphics[scale=0.35]{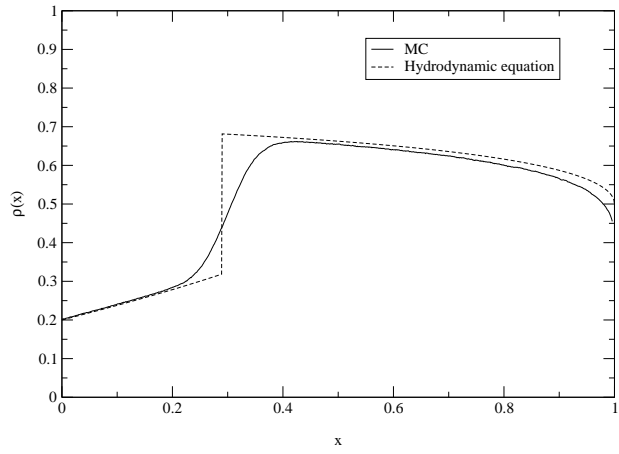}
\end{center}
\caption{
Plot of an average density of particles $\rho$ versus rescaled coordinate
$x$ ($site \ number/L$)
  of a localized density shock in the ASEP with 
Langmuir kinetics.
Parameters are $\rho_-=0.2$, $\rho_+=0.6$, $\Omega_a=0.3$ and $\Omega_d=0.1$.
We show the results of both Monte Carlo simulations for L=1000 and the mean field
approach.}
\label{ASEP-shock}
\end{figure}

One may equip the system with the
additional feature of local particle creation at empty sites with rate
$\omega_a$ and annihilation with rate $\omega_d$ (see
Fig.\ \ref{model})\cite{Parmeggiani,Challet}. 
In the thermodynamic limit $L \rightarrow \infty$ 
there are three regimes to be distinguished: If $\omega_a$ and
$\omega_d$ are of an order larger than $1/L$ the  
steady state of the system will be that of Langmuir kinetics,
i.e., there will be a uniform density of  
$K=\omega_a / (\omega_a+\omega_d)$ in the system. In case of
$\omega_a$ and $\omega_d$ being of smaller order than $1/L$, the local
kinetics is negligible and the system will behave 
as the TASEP. The case of the local rates being of order $1/L$ is the
most interesting one and will be investigated further on. 
Writing 
\begin{equation}
\label{omega}
  \omega_a=\Omega_a / L, \quad \omega_d=\Omega_d /L
\end{equation}
the phase diagram can be formulated in terms of
$\Omega_a$, $\Omega_d$, $\rho_-$ and $\rho_+$. In
Ref.\ \onlinecite{Parmeggiani} 
it was shown that for $\Omega_a$ and $\Omega_d$ fixed, the phase
diagram as a function of $\rho_-$ and $\rho_+$  
does not only exhibit the low-density and high-density phases known
from the TASEP, but also a high-low  
coexistence phase. In  this phase the shock does not move
in the system but its position is a function of the 
rates $\rho_-$ and $\rho_+$ (see Fig.\ \ref{ASEP-shock}). 

Parmeggiani et.al. presented not only Monte Carlo simulations but
derived also a mean field equation for the  
density profile which was shown to coincide with the simulation profiles.
We argue here that the mean field approximation can not be used
in general. The coincidence with the Monte Carlo in Ref.\
\onlinecite{Parmeggiani} is due to lack of correlations in true steady
state of the TASEP. 
We claim that the stationary density profile can be derived in general using a
hydrodynamic equation taking correlations into account (in case of the
TASEP this equation is equal to that obtained with a mean field
approach). For the Katz-Lebowitz-Spohn (KLS) model, which 
is a generic  model of interacting driven diffusive systems
\cite{KLS,Schmittmann} we show that this hydrodynamic equation correctly
describes the density profiles on a quantitative level, 
while a mean field approach would
fail to reproduce even basic qualitative features of the system, e.g.,
phase separation into three distinct density regimes.

\section{Hydrodynamic equation}
\label{Hydrodynamic}

%If not indicated otherwise we will use the rescaled variable $x=i/L$
%for indicating the position 
%of site $i$ in the lattice of size $L$. 

In the following we are interested in the $L\to\infty$ limit which we
perform by tuning the lattice spacing $a=1/L\to 0$ and rescaling of
time $t=t_\text{lattice}/L$ (Eulerian scaling) to get the continuous
(hydrodynamic) limit of the model. In this framework $\Omega_{a,d}$
are the attachment/detachment rates per unit length.
We claim that the
hydrodynamic equation describing the time dependence of the
local density $\rho(x)$ for a general driven diffusive system with Langmuir
kinetics takes the form
\begin{equation}
  \label{continuity} 
  \frac{\partial}{\partial t} \rho  + \frac{\partial}{\partial x}
  j (\rho) = {\cal L}(\rho), 
\end{equation}
where $j(\rho)$ is 
 the {\it exact} current in a driven diffusive system with homogeneous density
 $\rho$ without LK and ${\cal L}(\rho)$ 
the source term describing the Langmuir kinetics.
Here, we consider only that choice of ${\cal L}(\rho)$ which corresponds to
 the process depicted on Fig.\ \ref{model}:
\begin{equation}
\label{LK}
  {\cal L}(\rho) = \Omega_a (1-\rho(x,t)) - \Omega_d \rho(x,t)
\end{equation}
Other choices of ${\cal L}(\rho)$, which might e.g. describe the local
annihilation of particle pairs are 
to be discussed in a forthcoming publication \cite{coagulation}.

As is usually done in the rigorous derivation of the hydrodynamic limit of 
conservative systems \cite{Rezakhanlou91}, 
our nonconservative Eq.(\ref{continuity}) implicitly 
assumes that the system is locally stationary because the exact form of the 
stationary flux is used. We argue that this assumption is justified since the 
nonconservative part of the dynamics of the system at macroscopic scale
is so slow that locally the system reaches stationarity with respect to the
conservative part of the dynamics. Any finite perturbation caused by the 
nonconservative dynamics would travel a macroscopic distance and 
hence dissipate before interacting with another perturbation.
Hence the hydrodynamic description (after 
time rescaling $t\rightarrow \epsilon t$) is adequate for describing the
full dynamics. For physical insight in the formation of shocks 
one needs other tools which are discussed below.

Rewriting equation (\ref{continuity}) by using that $\partial_t
\rho(x,t)=0$ in the stationary 
 state and $\partial_x j = \partial j / \partial \rho \cdot \partial
 \rho / \partial x$ yields  
for the stationary density profile $\rho(x)$:
\begin{equation}
  \label{DE}
  v_c(\rho)\frac{ \partial \rho(x)}{ \partial x} = {\cal L}(\rho).
\end{equation}
Here, $v_c = \partial j / \partial \rho$ is the collective velocity,
i.e., the drift velocity of a center of mass 
of a local density perturbation  on a homogeneous stationary background 
with the density $\rho$ (for system with the Langmuir kinetics switched off)
 \cite{Habil,
  Kolomeisky}.
The stationary density profile has to satisfy (\ref{DE}) as
well as the boundary 
conditions $\rho(0)=\rho_-$ and $\rho(1)=\rho_+$. As equation
(\ref{DE}) is of the first order 
there will be in general no smooth solution fitting both boundary
conditions. In the original lattice model this discrepancy is resolved
by appearance of shocks and/or boundary layers. 
To regularize the
problem, one can add to (\ref{continuity}) and correspondingly to
  (\ref{DE})
 a vanishing viscosity term 
\begin{equation}
  \label{DE2}
  v_c (\rho)\frac{ \partial \rho(x)}{ \partial x} =  {\cal L}(\rho)+
  \nu \frac{ \partial^2 \rho(x)}{ \partial x^2}, 
\end{equation}
where $\nu>0$ is of order of $1/L$. 
This term makes the hydrodynamic equation second order
and ensures a smooth solution fitting both boundary conditions. The shock has 
then a width of order $1/L$ (see Ref.\ \onlinecite{Parmeggiani}), i.e., in the
thermodynamic limit the rescaled solution 
becomes discontinuous. We claim that equation (\ref{DE2}) gives the
same result in the $L\to\infty$ limit as the Monte Carlo, therefore it
can be used as a tool to compute the stationary density profile. The
main difference between (\ref{DE2}) and the MC is that the former does
not take fluctuations into account which leads to a shock width of
order $1/L$ while in a MC after averaging it is of the order of
$1/\sqrt{L}$ due to the fluctuation of the shock position.

The stationary density profile for a given $j(\rho)$ and parameters $\Omega_a$,
$\Omega_d$, $\rho_-$ and $\rho_+$ can be 
derived from the flow-field of the differential
equation (\ref{DE}) by using the rules, formulated and explained below:

\renewcommand{\labelenumi}{(\Alph{enumi})}
\begin{enumerate}
\item 
  In the interior of the lattice the stationary density profile
   either follows a line of the  flow field of the differential
  equation (\ref{DE}) or makes a jump. Jumps can only occur between
  densities yielding the same current, i.e., \textit{the current is
    continuous in the interior of the lattice}. 
\item
  Let $\rho_\pm'$ be defined as
limiting left and right densities with the boundary layers cut away:
  \[
  \rho_-' = \lim_{x\to +0} \rho(x), \quad \rho_+' =
  \lim_{x\to 1-0} \rho(x), 
  \]
  where $\rho(x)$ is the stationary profile in the hydrodynamic limit. 
  The boundary layer at $x=0$ ( i.e., if
  $\rho_-\neq\rho_-'$) has to satisfy the following condition: 
  \begin{eqnarray}
\label{B1}
  \text{if } \rho_-<\rho_-' \text{ then } j(\rho)>j(\rho_-') \text{ for
    any } \rho\in(\rho_-,\rho_-') \\
\label{B2}
  \text{if } \rho_->\rho_-' \text{ then } j(\rho)<j(\rho_-') \text{ for
    any } \rho\in(\rho_-',\rho_-)
  \end{eqnarray}
  The condition for the stability of the boundary layer at $x=1$ (if there is) is
  similar:
  \begin{eqnarray}
\label{B3}
  \text{if } \rho_+'<\rho_+ \text{ then } j(\rho_+')<j(\rho) \text{ for
    any } \rho\in(\rho_+',\rho_+) \\
\label{B4}
  \text{if } \rho_+'>\rho_+ \text{ then } j(\rho_+')>j(\rho) \text{ for
    any } \rho\in(\rho_+,\rho_+')
  \end{eqnarray}
\item
  Shocks between a density $\rho_l$ to the left of the shock and
  $\rho_r$ to the right of the shock are stable only if they are stable in the
  absence of Langmuir kinetics\cite{Habil,Hager}. 
\end{enumerate}
\renewcommand{\labelenumi}{\arabic{enumi}.}
Remarks:
\begin{itemize}
\item 
Although LK does not conserve locally the number of particles,
Eq. (\ref{continuity}) with the vanishing viscosity 
added (\ref{DE2}) can be rewritten formally in the form
\begin{equation}
  \label{cons_law} 
  \frac{\partial\rho (x,t) }{\partial t}  + \frac{\partial}{\partial x}
  \tilde j (x,t) = 0,
\ \ 
 \tilde j (x,t) = j(\rho)- \int_A^x {\cal L}(\rho) dx - 
\nu \frac{\partial \rho}{\partial x} - {\cal F}(t)
\end{equation}
where ${\cal F}(t)$ is some time-dependent function. Suppose that there is 
a shock at the position $X_0$ connecting the densities $\rho_l$ and 
$\rho_r$. The mass transfer across the shock is
\begin{equation}
  \label{mass_transfer} 
 \frac{\partial }{\partial t}\int_{X_0-0}^{X_0+0}
 \rho (x,t)  dx = 
\tilde j (X_0+0,t)-\tilde j (X_0-0,t) = j (\rho_r)-j (\rho_l),
\end{equation}
since the Langmuir term and the viscosity term change only infinitesimally
across the shock. In the stationary state, the RHS of (\ref{mass_transfer}) 
vanishes which explains the rule (A).

\item
The rule (B) is due to the fact that in the boundary layer 
of the vanishing length $\delta l \rightarrow 0$, the LK term
in (\ref{cons_law}) can be neglected. Consequently, for the 
stationary current at the boundaries we have 
$\tilde j(x) = j(\rho(x))- \nu  \frac{\partial\rho }{\partial x}=J$,
which yields the known  
maximization/minimization principle\cite{Habil,Popkov}, and is equivalent 
to rule (B). Indeed at the left boundary  
$J=j(\rho_-')$ (see (\ref{B1}) for  notations), and if, e.g., 
$ \rho_- <\rho_-'$, then
$\frac{\partial\rho }{\partial x}>0$. Consequently,
we obtain $j(\rho_-)= J + \nu\frac{\partial\rho }{\partial x} > J$,
which is exactly  (\ref{B1}). Analogously one obtains 
(\ref{B2})-(\ref{B4}).

%  Condition (B) can be applied without Langmuir kinetics too. In
%  this case the solution of (\ref{DE}) is $\rho(x)=$ const., which
%  means that $\rho_-'=\rho_+'=\rho_{\text{bulk}}$ and we get the known
%  current maximization/minimization principle\cite{Habil,Hager}.
\item
The rule (C) is explained by the marginal role the Langmuir kinetics
plays locally in space and in time. 
The first, LK is very slow locally for large $L$ 
(see (\ref{omega})), and the second,
it acts ``orthogonally'' on the particle distribution,
not affecting directly the particle motion. Hence, the local 
perturbations will still spread with the velocity corresponding
to the local density level $\rho$, thus rendering the same
stability conditions for a shock  as for the diffusive system 
without LK.

Condition (C) is easy to check geometrically through the current-density relation: an
upward (downward) shock is stable if the straight line connecting
the points $(\rho_l,j(\rho_l))$ and $(\rho_r,j(\rho_r))$ stays below (above)
the $j(\rho)$ curve\cite{Hager,Popkov}. Because of criterion (A) these lines are always
horizontal in this case which gives zero mean velocity (but not
localization) for the shock in absence of Langmuir kinetics.
\item
In the cases we have considered (ASEP, KLS model), the rules (A)-(C) 
define an unique stable solution (see an Appendix) and we believe
that this is true also in general case, i.e., for arbitrary
$j(\rho)$ dependence and given choice (\ref{LK}) of Langmuir kinetics.
\end{itemize}

In the following we apply the general theory to specific models. 

\section{Revisiting the ASEP  with Langmuir kinetics}

Using the differential equation (\ref{DE}) and the rules given above we reconsider
the  ASEP  with Langmuir kinetics \cite{Challet,Parmeggiani}.
 Here, the current-density relation is given by 
$j( \rho)=\rho (1- \rho)$, which yields $v_c (\rho) = 1-2 \rho$. Thus
equation (\ref{DE}) becomes 
\begin{equation}
  \big(1-2 \rho(x)\big)\partial_x \rho(x) = \Omega_a - ( \Omega_a + \Omega_d) \rho(x),
\end{equation}
which is identical with the mean field equation in Ref.\ \onlinecite{Parmeggiani} in the
thermodynamic limit. We would like to stress that this coincidence is caused by the fact
that the mean field current-density relation for the TASEP is
exact. As is demonstrated below, 
equation (\ref{DE}) also holds when this is not the case,
as e.g. for the one-dimensional 
KLS model.

Due to rule (A) as stated above (continuity of the current in the
interior of the lattice)  
shocks in the interior can only occur in the case that $\rho_l = 1-
\rho_r$, as $j ( \rho)$ is 
symmetric to $\rho = 1/2$. Rule (C) (stability of the shock) furthermore requires that 
$\rho_r > \rho_l$. These observations coincide with 
 the findings of \cite{Parmeggiani}.

We also applied our rules to 
$k$-hop exclusion models \cite{Binder} ( with LK added), which are a generalization
of the TASEP with stationary product measures
and asymmetric current-density relations. Due to 
this fact shocks appear, which are non-symmetric with respect to 
$\rho=1/2$. MC simulations are in full accord with our predictions 
\cite{Willmann_unpublished}.

\section{KLS model with Langmuir kinetics}
\label{KLS_model}

A much studied one-dimensional driven diffusive system with
interactions between the particles 
is the following variant of the KLS model \cite{Mukamel2,Hager,Popkov}:

In the interior, particles at site $i$ move to site $i+1$, provided it
is empty, with a rate that 
depends on the state of sites $i-1$ and $i+2$.

\begin{eqnarray*}
0100 \rightarrow 0010 & \text{ with rate } & 1+\delta \\
1100 \rightarrow 1010 & \text{ with rate } & 1+\epsilon \\
0101 \rightarrow 0011 & \text{ with rate } & 1-\epsilon \\
1101 \rightarrow 1011 & \text{ with rate } & 1-\delta \\
\end{eqnarray*}

\begin{figure}[h]
\begin{center}
\includegraphics[scale=0.30]{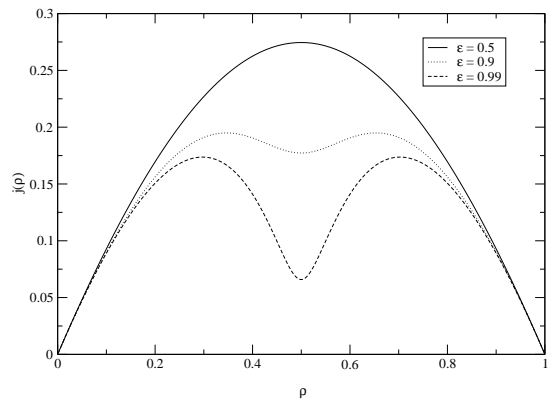}
\end{center}
\caption{ Current-density relation for the one-dimensional KLS model for various 
$\epsilon$. }
\label{current-density-KLS}
\end{figure}

At site 1 particles can enter the lattice provided the target site is
empty. The rate depends on 
the state of site 2. Similarly, particles can leave the system at
site $L$ with a rate depending 
on the state of site $L-1$. The boundaries mimic the action of
reservoirs with densities  
$\rho_-$ and $\rho_+$. For $\rho_-=\rho_+$ the stationary state is
that of an one-dimensional  
Ising model with boundary fields. The current-density relation can be
calculated exactly using 
transfer matrix techniques \cite{Hager}. It turns out that for strong
enough repulsion between 
the particles ($\epsilon  \gtrsim 0.9$) a current-density relation with two
maxima arises (see Fig.\ \ref{current-density-KLS}). 
The parameter $\delta$ determines the skewness of $j(\rho)$ with
respect to the vertical 
line  $\rho=1/2$. For 
$\delta=0$, the system has particle-hole symmetry resulting in
 $j(\rho)$ being symmetric with respect to 1/2. For simplicity we
consider this case in the rest of the paper.

The phase diagram of this family of models with strong particle
repulsion is known to exhibit  
7 different phases, among them two maximal-current and one minimal-current phase. 
The phase diagram is determined by the interplay of diffusion, branching and
coalescence of shocks \cite{Popkov}.

When equipping these models with Langmuir kinetics one expects that a
very rich phase diagram 
with many more than the original 7 phases will appear. We will not
attempt to give this full 
phase diagram here, but instead present two new features, which cannot
be observed in systems 
without a concave region in the
%single-maximum 
current-density relation: localized downward
shocks and double shocks. 

\subsection{Localized downward shocks}

\begin{figure}[h]
\vspace{1cm}
\begin{center}
\includegraphics[scale=0.35]{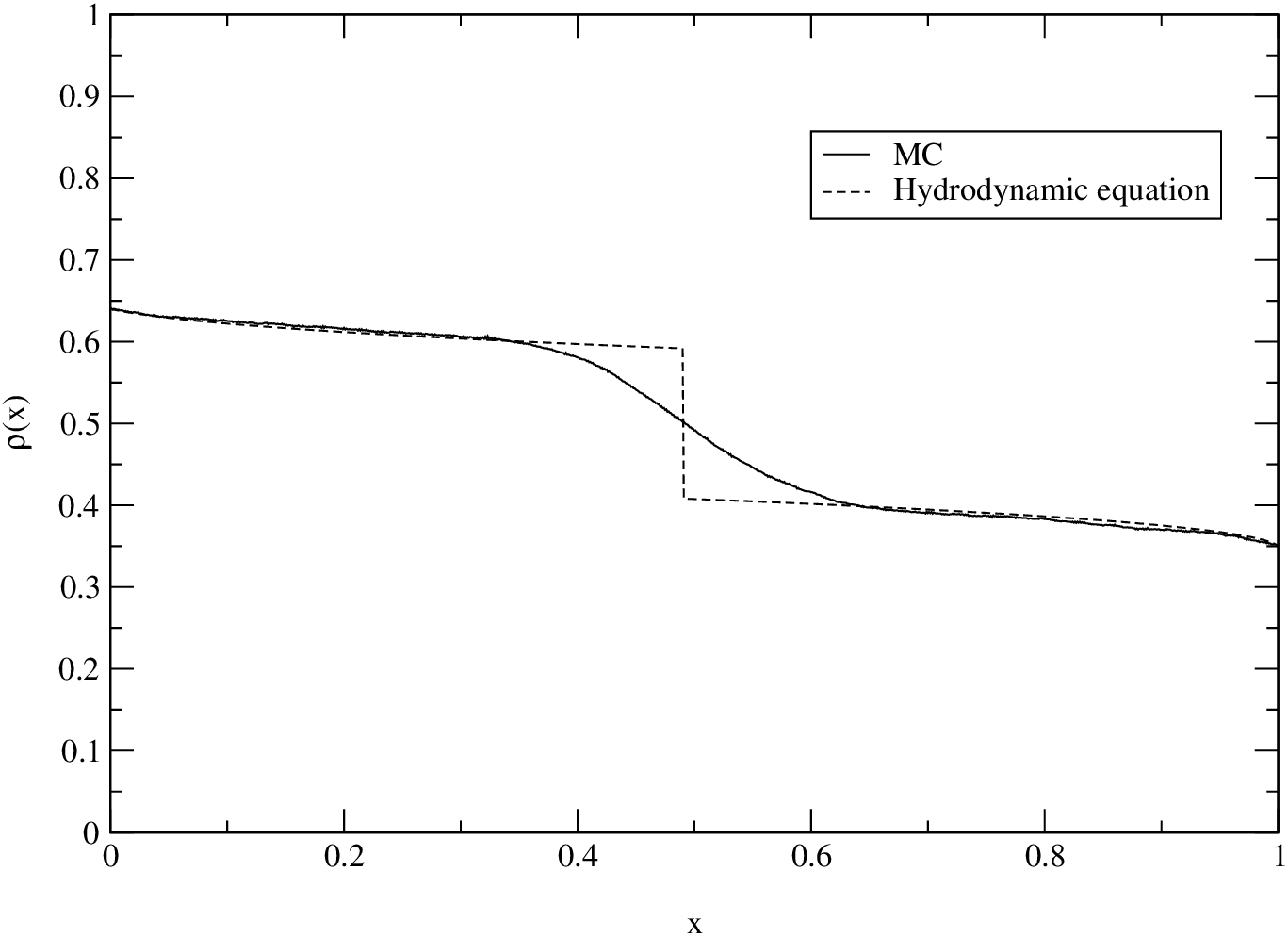}
\end{center}
\caption{Density of particles $\rho$ versus rescaled coordinate
$x$ ($site \ number/L$)
 in a localized downward shock in the KLS model with Langmuir
kinetics. Parameters are $\rho_-=0.64$, $\rho_+=0.35$, $\Omega_a=\Omega_d=0.05$.
We show the results of both hydrodynamic equation and Monte Carlo simulation 
for $L=1000$. The smoothness of the 
MC result is due to the fluctuation of the shock position\cite{coagulation}.}
\label{downward-shock}
\end{figure}

In the regime where the current-density relation of the KLS model exhibits two
maxima at densities $\rho_1^*$ and $\rho_2^*$, where $\rho_1^*<\rho_2^*$ and a
minimum at $\rho=1/2$ (at $\delta=0$) there is a region where downward
shocks are stable according to Ref.\ \onlinecite{Hager,Popkov} (and
rule (C)). These
are characterized by $\rho_l\in(0.5,\rho_2^*)$ and
$\rho_r\in(\rho_1^*,0.5)$. This suggests that localized downwards
shocks may appear when introducing the kinetic rates. 
%For $\rho_- \subset (0.5,\rho_2^*)$ and
%$\rho_+ \subset (\rho_1^*,0.5)$ a downward shock can form, see Fig.~
%\ref{downward-shock}.
In deed, in the KLS model with Langmuir kinetics
for certain values of the boundary densities $\rho_-$ and $\rho_+$
,which strongly depend on the kinetic rates $\Omega_a$ and
$\Omega_b$, one gets a stable downward shock according to rules
(A,B,C). We give an example for this case on Fig.\ \ref{downward-shock}. 

One can see that employing the general theory described above yields a
stationary profile with a localized downward shock, which coincides
with the MC results up to finite size effects, while a simple
mean field approach would fail as it would not be able to capture the
difference between the KLS model with $\epsilon>0$ and the TASEP (KLS
with $\epsilon=0$). 

%For a profile as exemplified by 
%Fig.~7, the shock velocity is 0, while $v_c(\rho_l)>0$ and $v_c(\rho_r)<0$
%(see Fig.~ \ref{downward-shock-2}). Thus, the stability criterion (rule (C)) is
%satisfied.
%Models which lack a concave region in the current-density relation do
%not show stable downward shocks, due to rule (C). 

\subsection{Localized double shocks}
\label{B}

%As is known from the studies of the KLS model \cite{Popkov,Hager}, for $\rho_-<\rho_+$
%the system will settle into a steady state which minimizes the current. In case of
%the KLS model with LK this can for certain boundary conditions be achieved
%by bridging the double maxima structure of the current-density
%relation with two subsequent shocks (see Fig.\ \ref{double-shock} and \ref{double-shock-2}). 
%Let
%$\tilde {\rho_1}$ and $\tilde {\rho_2}$ be defined as the two
%densities which yield 
%the same current as the local minimum $\rho=1/2$ and $\tilde
%{\rho_1}<\tilde {\rho_2}$. If 
%$\rho_-<\tilde {\rho_1}$ and $\rho_+>\rho_2^*$ a double shock can
%arise.
%The first shock
%between the densities $\rho_{l1}=\tilde {\rho_1}$ and $\rho_{r1}=1/2$ bridges the first 
%maximum in the current-density relation. According to rule (C) this shock is stable.
%The second shock between $\rho_{l2}>1/2$ and $\rho_{r2}<\rho_+$ bridges the second
%maximum and is also stable.

%Also here Monte Carlo simulations and the hydrodynamic equation (\ref{DE2}) coincide,
%as can be seen in Fig.\ \ref{double-shock}. A simple mean field approach could not predict a double
%shock. 
%Note that for a hypothetical current $j(\rho)$
%containing 3 and more maxima, triple etc. localized shock can be constructed 
%following the same strategy.

Let $\rho_{1,2}'$ be defined as the inflection points of the
current-density relation ($\rho_1'<\rho_2'$).
As is known from the studies of the KLS model \cite{Popkov,Hager}, if
we start an infinite system from a step-like initial density profile
with $\rho_-\in (\tilde\rho_1,\rho_1')$ on the left and $\rho_+\in
(\rho_2',\tilde\rho_2)$ on the right, we get a time-dependent
solution having two shocks: One of these has negative mean velocity,
while the other has positive and in the middle there is an expanding
region with $\rho=1/2$ (for $\delta=0$) which corresponds to the
minimal current phase in a system with open boundaries\cite{Hager,Popkov}.

\begin{figure}[h]
\begin{center}
\includegraphics[scale=0.35]{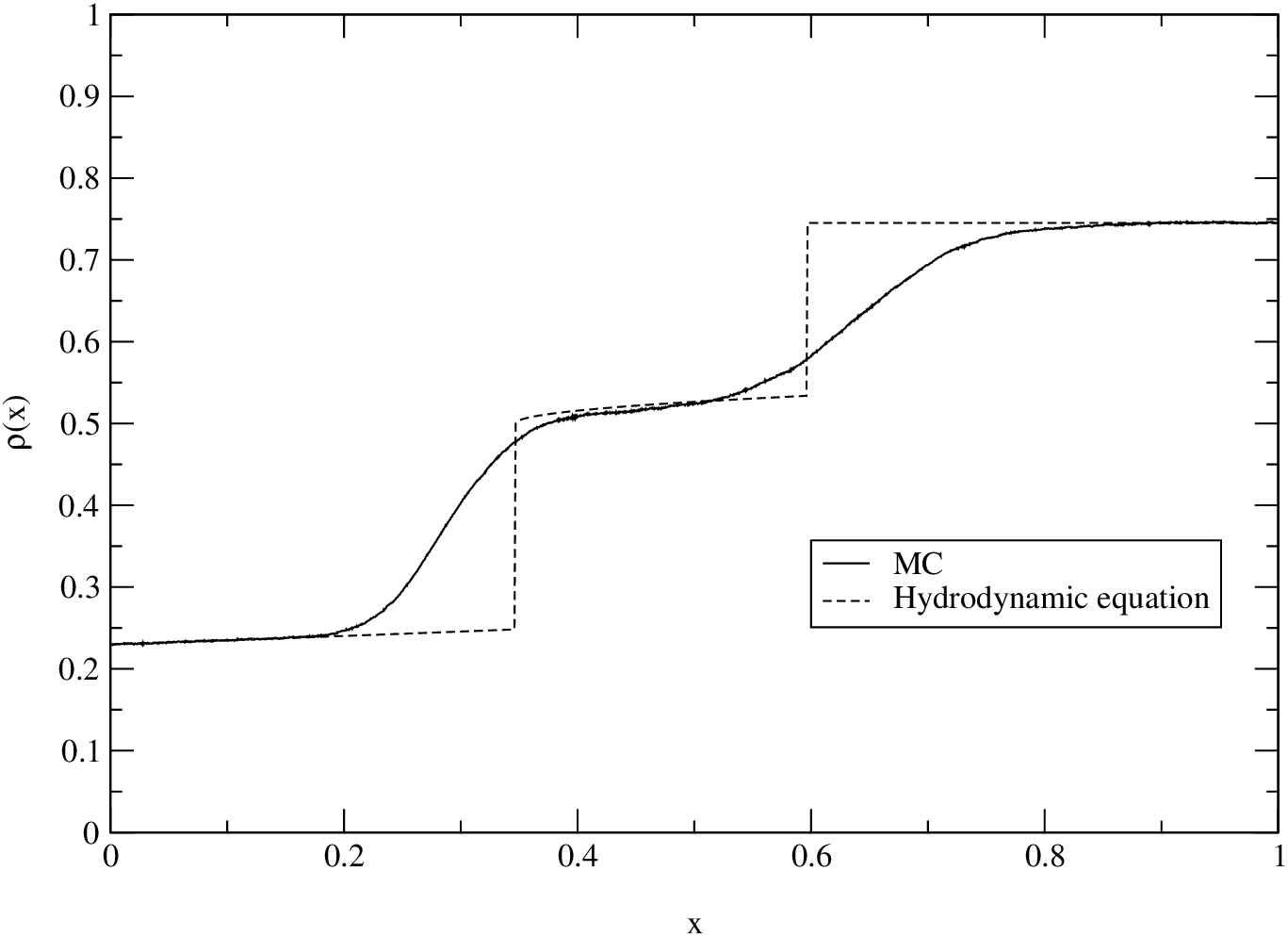}
\end{center}
\caption{Density of particles $\rho$ versus rescaled coordinate
$x$ ($site \ number/L$) 
in a localized double shock in the KLS model with Langmuir kinetics.
Parameters are $\rho_-=0.23$, $\rho_+=0.745$, $\Omega_a=0.03$ and $\Omega_d=0.01$.
We show the results of both hydrodynamic equation and Monte Carlo simulation 
for $L=1000$. The smoothness of the 
MC result is due to the fluctuation of the shock position\cite{coagulation}.}
\label{double-shock}
\end{figure}

This leads us to the conjecture that introducing the kinetic rates for
certain values of $\rho_-,\rho_+, \Omega_a, \Omega_d$ one may
achieve a stable double shock structure. In Fig.\ \ref{double-shock} we
present an example for such a case. Application of rules (A,B,C), which is
presented in detail in Appendix \ref{A}, yields
the same double shock structure as the MC up to finite size effects. Note, that
a simple mean field approach could not predict a double shock.

\section{Conclusions}
 
In this work we present a hydrodynamic equation which, together with
some rules treating the discontinuities, correctly describes the 
stationary states of one-dimensional driven diffusive systems with Langmuir kinetics
and open boundaries. It captures both systems without correlations
in a steady state ( as e.g. the 
TASEP and the $k$-hop exclusion models) and systems with correlations as the KLS model.
For the latter the two new phenomena of a stationary localized downward shock and
a localized double shock (corresponding to phase separation to three
distinct regions) were presented which a mean field approach would not 
reproduce. The exact currect of driven diffusive systems without LK
enters the hydrodynamic description since the bulk has sufficient time
to relax between subsequent annihilation/creation events.
 An interesting, paradoxical feature of
these phenomena is that fluctuating shocks get localized due to extra
noise (LK), which is highly unexpected.

\begin{acknowledgments}

We wish to thank the authors of Ref.~\onlinecite{Parmeggiani} for communicating
their results prior to publication and M. Salerno for useful discussions. 
AR acknowledges financial support by Deutsche Forschungsgemeinschaft.
ABK acknowledges the support of the Camille and Henry Dreyfus New Faculty
Awards Program (under Grant NF-00-056)  and the hospitality of
Forschungszentrum Julich.
\end{acknowledgments}

\appendix*
\section{Double shock density profile from the rules (A)- (C)
 }
\label{A}

Here we  demonstrate how one  determines the stationary
density profile using the rules (A), (B) and (C) from the
section~\ref{Hydrodynamic}. As an example we
take the parameters which yield  a double (localized) shock
structure in the KLS model ($\rho_-=0.23$, $\rho_+=0.745$,
$\Omega_a=0.03$ and $\Omega_d=0.01$). The KLS-model parameters are:
$\delta=0, \epsilon=0.9$ (see  section~\ref{KLS_model}).

\begin{figure}[h]
\begin{center}
\includegraphics[scale=0.7]{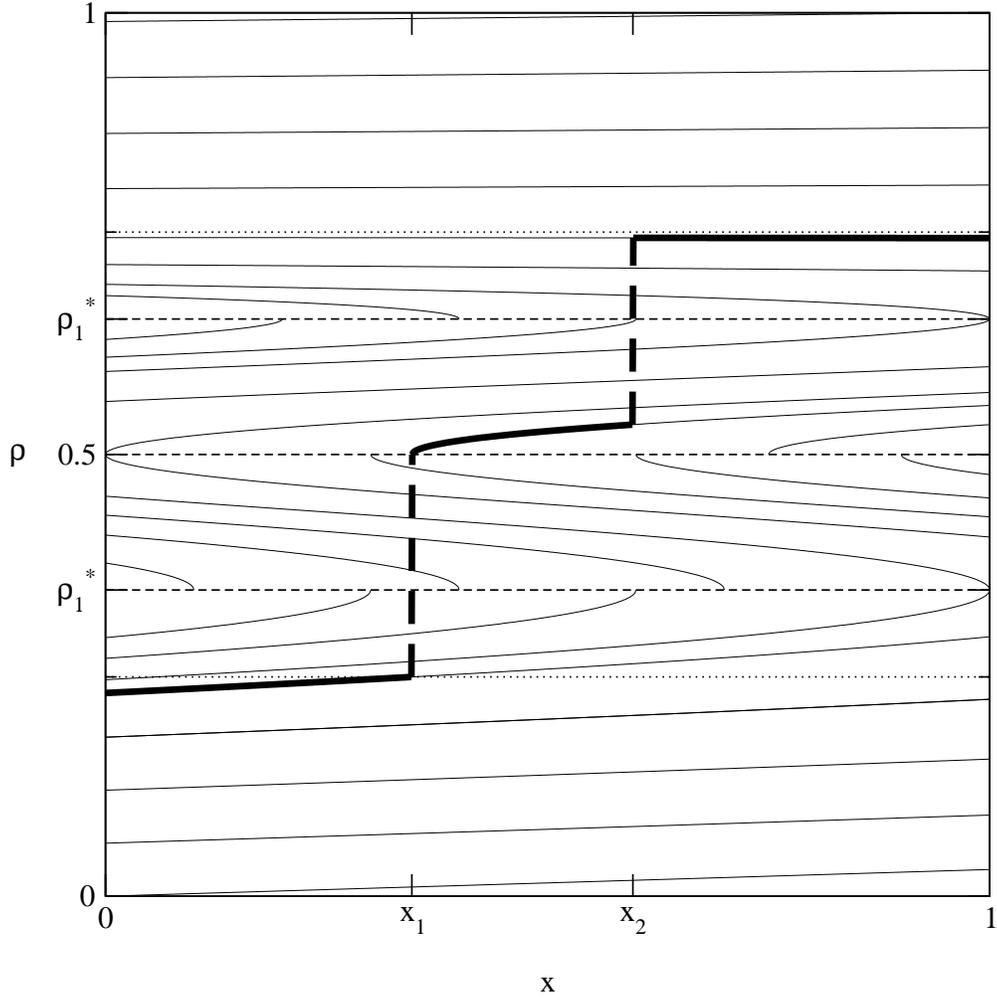}
\end{center}
\caption{The flowfield of the hydrodynamic equation in the KLS model
  with Langmuir kinetics. Parameters are $\delta=0, \epsilon=0.9, \Omega_a=0.03,
  \Omega_d=0.01$. The thick lines show the stationary density profile
  for $\rho_-=0.23, \rho_+=0.745$ given by the rules (A,B,C).
The dotted lines are
$\rho=\tilde \rho_1\approx 0.24821$, $\rho=\tilde \rho_2\approx  0.75178$
(see the subsection~\ref{B} for notations).
 The axes: $x$ is a rescaled coordinate ($site \ number/L$), $\rho(x)$ is an
  average density of particles at point $x$.
}
\label{flowfield}
\end{figure}

First suppose that there is a boundary layer at $x=0$. According to
rule (B) it is stable only if $\rho_-'>1-\rho_-=0.77$. If this is the
case then in the bulk there is no allowed jump 
since these trajectories of the flow-field (see Fig.\ \ref{flowfield}) stay
always above $\rho=0.75$ (rules (A) and (C)) which yields
$\rho_+'>0.75$. But then  the boundary layer at $x=1$ does
not satisfy rule (B). This contradiction shows that
 there is no boundary layer at
$x=0$. One can use the same argument to show that there is no boundary
layer at $x=1$ either.

Now one can see that the stationary density profile close to the left
boundary follows the line of the flow-field for which
$\rho(x=0)=\rho_-=0.23$. Since there is no boundary layer at the right end it
is clear that somewhere in the bulk it has to make a jump. 

Note that this trajectory crosses the
line $\rho=\tilde\rho_1$ at $x=x_1$. 
Suppose that the jump takes
place before at $x<x_1$. In this case, according to rule (A)
it would jump over $\tilde\rho_2=1-\tilde\rho_1$ which would result in
a boundary layer at $x=1$ which is not allowed. If the jump takes place
at $x>x_1$ then 
$\rho_1^*<\rho_r<0.5$ and since from this region there is no
allowed jump it would end up at $\rho_1^*<\rho_+'<0.5$
resulting again in an unstable boundary layer on the right side. This shows
that the jump is located at $x=x_1$ and from here the density profile
follows the trajectory which starts at $x=x_1$ with the value
$\rho=0.5+0$. 

One can easily see that we need another jump to connect this
trajectory with the one which ends at $x=1$ with
$\rho=\rho_+$. Applying rule (A) (continuity of the current) we can
get the point $x_2$ where the second jump is located.

%\newpage
\vspace{1cm}
%\newpage
%\newpage
%\newpage
%\newpage
\begin{figure}[h]
\vspace{0.5cm}
\begin{center}
\includegraphics[scale=0.35]{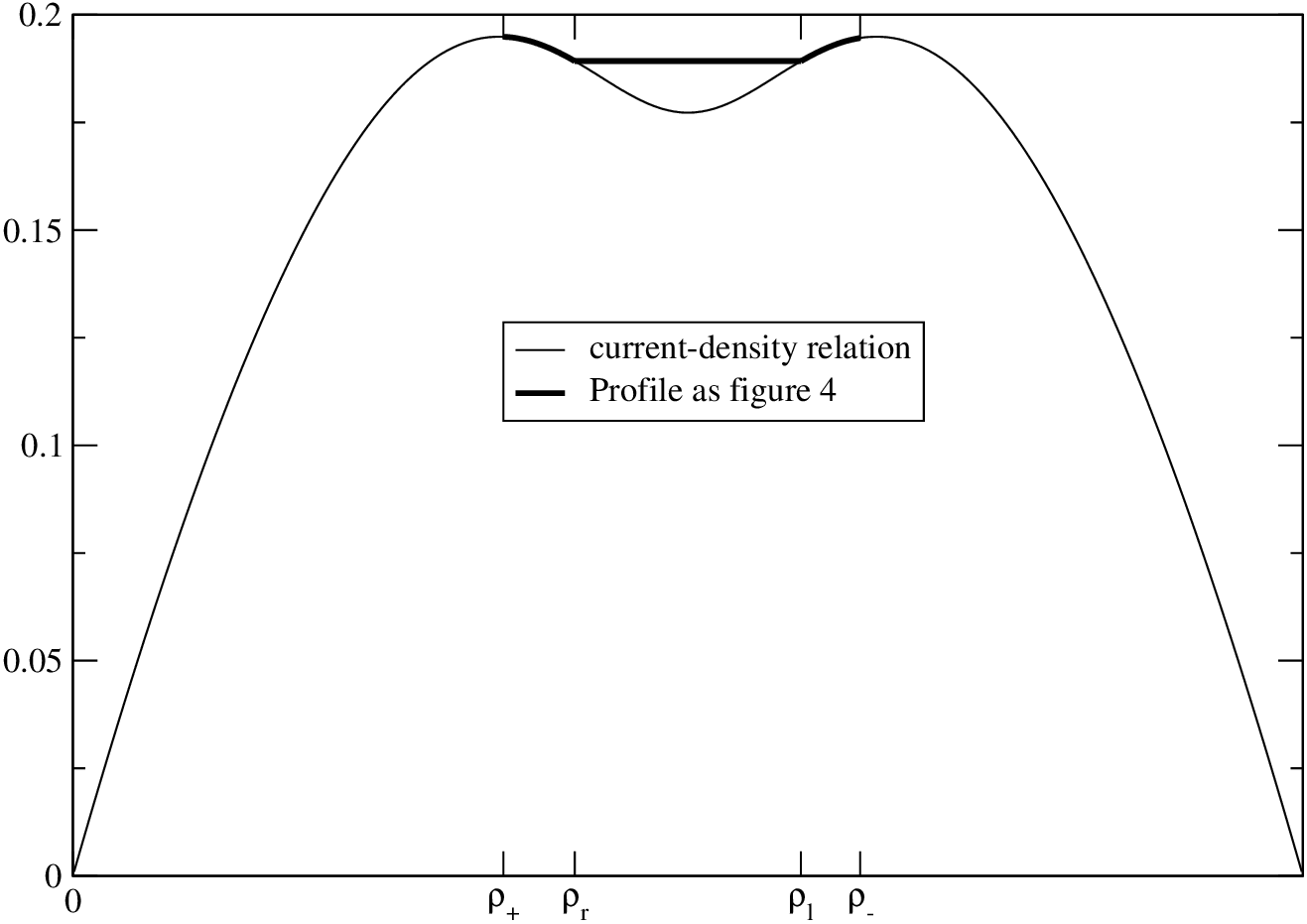}
\end{center}
\caption{Path in the current-density relation for the profile shown in figure \ref{downward-shock}.}
\label{downward-shock-2}
\end{figure}
%\newpage
%\newpage
\begin{figure}[h]
\begin{center}
\includegraphics[scale=0.35]{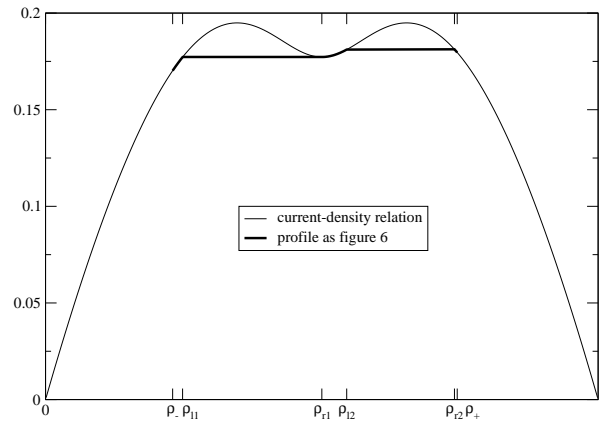}
\end{center}
\caption{Path in the current-density relation for the profile shown in figure \ref{double-shock}.
}
\label{double-shock-2}
\end{figure}
%\newpage

\begin{thebibliography}{99}
\bibitem{Habil} G. M. Sch\"utz in \textit{Phase Transitions and Critical Phenomena} 
Vol 19, Eds. C. Domb and J. Lebowitz (Academic, London, 2000).
\bibitem{Krug} J. Krug, Phys. Rev. Lett. \textbf{67}, 1882 (1992).
\bibitem{Evans} M. R. Evans, Y. Kafri, H. M. Koduvely, and D. Mukamel, Phys. Rev. Lett. \textbf{80}, 425 (1998).
\bibitem{Mukamel95}  Evans  M R,  Foster D P, Godr\`eche C and
 Mukamel D,
 J. Stat. Phys. {\bf 80}  69 -102 (1995)
\bibitem{Mukamel1}Y. Kafri, E. Levine, D. Mukamel, G. M. Sch\"utz and J. T\"or\"ok, Phys. Rev. Lett. \textbf{89}, 035702 (2002).
\bibitem{Mukamel2}Y. Kafri, E. Levine, D. Mukamel, G. M. Sch\"utz and R.D. Willmann, \textit{cond-mat/0211269}.
\bibitem{Kolomeisky2} N. Mirin and A. B. Kolomeisky, J. Stat. Phys. \textbf{110}, 811 
(2003).
\bibitem{Parmeggiani} A. Parmeggiani, T. Franosch and E. Frey,
Phys. Rev. Lett. {\bf 90}, 086601 (2003).

\bibitem{Ligg75}
T.M. Liggett, Trans. Amer. Math. Soc. {\bf 179}, 433 (1975).
\bibitem{Domany} G. Sch\"utz and E. Domany, J. Stat. Phys. \textbf{72}, 277 (1993).
\bibitem{ASEP} 
Derrida B,  Evans M R,  Hakim  V and  Pasquier V  1993,
 J.Phys.A {\bf 26}  1493 (1993) ; 

\bibitem{Kolomeisky} A. Kolomeisky, G. M. Sch\"utz, E. B. Kolomeisky and J. P. Straley, J. Phys. A \textbf{31}, 6911 (1998).
\bibitem{Challet} R. D. Willmann, G. M. Sch\"utz and D. Challet, Physica A \textit{316}, 430 (2002).
\bibitem{KLS} S. Katz, J. L. Lebowitz and H. Spohn, J. Stat. Phys. \textit{34}, 497 (1984).
\bibitem{Schmittmann} B. Schmittmann and R.K.P. Zia, in \textit{Phase Transitions and Critical Phenomena}, Vol.17, Eds. C. Domb and J. Lebowitz (Academic Press, London 1995).
\bibitem{coagulation} A. R\'akos, M. Paessens 
and G. M. Sch\"utz, in preparation.

\bibitem{Rezakhanlou91} F. Rezakhanlou, Comm. Math. Phys.
{\bf 140}, 417 (1991).


\bibitem{Hager} J. S. Hager, J. Krug, V. Popkov and G. M. Sch\"utz, Phys. Rev. E \textbf{63}, 
056110 (2001).
\bibitem{Binder} P.-M. Binder, M. Paczuski and M. Barma, Phys. Rev. E \textbf{49}, 1174 (1994).
\bibitem{Willmann_unpublished} R. D. Willmann, unpublished
\bibitem{Popkov} V. Popkov and G. M. Sch\"utz,
  Europhys. Lett. \textbf{48}, 257 (1999).
%\bibitem{motorprot1} K. Kruse and K. Sekimoto, Phys. Rev. E {\bf 66},
%  031904 (2002).
%\bibitem{motorprot2} M. J. Schnitzer, K. Visscher and S.M. Block,
%  Nature Cell Biology {\bf 2}, 718 (2000)
%\bibitem{motorprot3} M. E. Fisher and A. B. Kolomeisky, Proc. Natl
%  Acad. Sci USA {\bf 98}, 7748 (2001)
\bibitem{motorprot4} A. Alberts {\em et al., The Molecular Biology of
    the Cell }, (Garland, New York, 1994)
\end{thebibliography}
\end{document}